\documentclass[10pt,letterpaper]{article}
\usepackage{opex3}

\usepackage{amsmath,cite}
\usepackage[colorlinks=true,breaklinks=true]{hyperref}

\begin{document}

\title{
Soliton compression to few-cycle pulses with a high quality factor by engineering cascaded quadratic nonlinearities
}

\author{Xianglong Zeng,$^{1,2}$ Hairun Guo,$^1$ Binbin Zhou,$^1$ and Morten Bache$^{1,*}$}
\address{$^1$Ultrafast Nonlinear Optics group, DTU Fotonik, Department of Photonics Engineering,
Technical University of Denmark, DK-2800 Kgs. Lyngby, Denmark\\
$^2$The Key Lab of Specialty Fiber Optics and Optical Access Network, Shanghai University, 200072 Shanghai, China}

\email{*moba@fotonik.dtu.dk} 



\begin{abstract}
We propose an efficient approach to improve few-cycle soliton compression with cascaded quadratic nonlinearities by using an engineered multi-section structure of the nonlinear crystal. By exploiting engineering of the cascaded quadratic nonlinearities, in each section soliton compression with a low effective order is realized, and high-quality few-cycle pulses with large compression factors are feasible. Each subsequent section is designed so that the compressed pulse exiting the previous section experiences an overall effective self-defocusing cubic nonlinearity corresponding to a modest soliton order, which is kept larger than unity to ensure further compression. This is done by increasing the cascaded quadratic nonlinearity in the new section with an engineered reduced residual phase mismatch. The low soliton orders in each section ensure excellent pulse quality and high efficiency. Numerical results show that compressed pulses with less than three-cycle duration can be achieved even when the compression factor is very large, and in contrast to standard soliton compression, these compressed pulses have minimal pedestal and high quality factor.
\end{abstract}



\section{Introduction}

High-energy pulsed lasers by employing chirped-pulse-amplification (CPA) technique are developed rapidly both in bulk solid-state and fiber amplifiers. Recently a power scaling approach by using coherent combination of several channels of femtosecond fiber CPA systems was demonstrated to hugely increase the deliverable power \cite{Seise2011}. Besides the popular Ti:sapphire technology, currently there is a lot of focus on CPAs at the wavelength of 1.03 $\mu{\rm m}$ by using Yb-doped or 1.56 $\mu{\rm m}$ by using Er-doped gain materials. However, their pulse durations are severely limited to several hundreds of femtosecond due to the narrow gain bandwidth of the active materials. Generation of optical few-cycle pulses inside the CPA systems requires nonlinear phase shift to broaden the spectrum, which may lead to the distorted pulses due to excessive detrimental nonlinear effects.

Pulse compression to few-cycle duration from hundreds of femtosecond duration out of the energetic solid-state and fiber amplifiers, would give a stable and compact optical source for e.g. ultrafast pump-probe spectroscopy and biology. One way of achieving this would be to use gas-filled cells \cite{Nisoli1996} or hollow capillaries \cite{Hauri2004} to achieve spectral broadening followed by dispersive optics like prisms or gratings to compress the pulses temporally, but these techniques, while standard, suffer from being hard to control and being very bulky.

An alternative is to use soliton compression from self-defocusing cascaded nonlinearities (e.g., phase-mismatched second-harmonic generation, SHG): when they are strong enough to overcome the inherent cubic (Kerr) self-focusing nonlinearity in the nonlinear materials, an effective self-defocusing nonlinearity is formed \cite{liu1999} without beam filamentation or break-up of the transverse part of the beam, problems often encountered with self-focusing nonlinearities, and the compressible pulse energy is in principle only limited by the aperture of the nonlinear medium. Furthermore solitons can be generated in the visible and near-IR regimes \cite{Ashihara2002}. This is possible because the effective negative nonlinear refractive index can be counterbalanced by the positive (normal) dispersion to create a soliton. When taking the effective soliton order from the total defocusing nonlinearity substantially larger than unity, in the initial propagation stage the input pulse will experience a dramatic self-compression effect. This soliton compression works in the same way as standard soliton compression with Kerr nonlinearities in fibers. Soliton compression by using cascaded quadratic nonlinearities was experimentally observed from 120 fs down to around 30 fs at the wavelength of 800 nm \cite{Ashihara2002} and at telecom wavelengths \cite{Ashihara2004,zeng2008}. 12 fs (3 optical cycles) compressed from 110 fs was achieved at 1250 nm \cite{Moses2007OL}. Multi-stage pulse compression in sequence by use of cascaded quadratic nonlinearity was proposed to achieve extreme compression factor from several ps to 30 fs \cite{Xie2007OptComm}. Quasi-phase-matching (QPM) technique offer additional degrees of freedom controlling cascaded nonlinearities, depending on the effective phase mismatch. Through engineering the QPM structure, adiabatic and high-quality pulse formation of quadratic soliton was presented \cite{ZengOE2006,ZengOE2009}.

Recently a temporal nonlocal model gave insight in optimal phase-mismatch ranges, where few-cycle pulse durations are possible \cite{BacheOL2007} and this led to an experimental demonstration of 17 fs compressed soliton (less than 4 optical cycles) from 47 fs near-IR pulses through cascaded quadratic soliton compression (CQSC) in a short standard lithium niobate crystal \cite{ZhouPRL}.

\begin{figure}[htbp]
  \centering{

  \includegraphics[width=10cm]{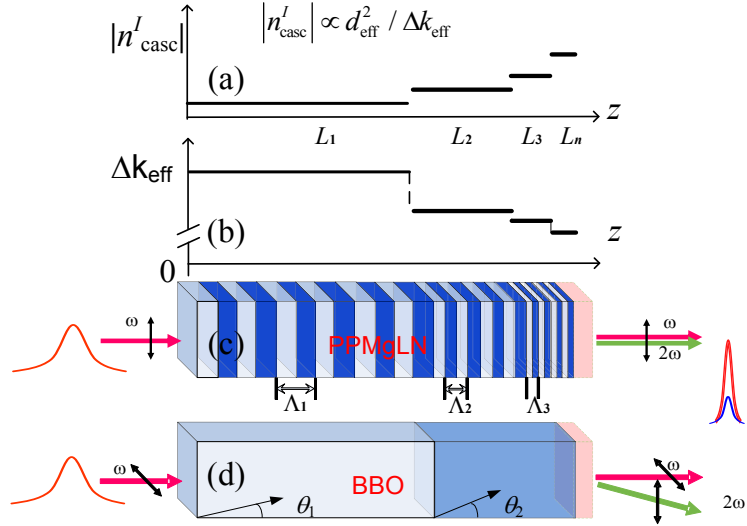}
}
\vspace{-2mm}
\caption{A schematic of multi-section structure of nonlinear crystals. (a) Cascaded nonlinearity increases and (b) the effective phase mismatch deceases upon propagation by controlling the local domain period $\Lambda_j$ in (c) QPM grating or the angle $\theta_j$ between the FW input and the optical $\textit{z}$ axis of the crystal in (d) bonded BBO crystals.}
\label{fig1}
\end{figure}

However, the soliton compression mechanism used in the connection with cascaded nonlinearities has similar drawbacks to the higher-order soliton compression known from optical fibers: the larger the soliton number, the larger the unwanted pedestal, and the less energy remains in the compressed spike. Such a feature is detrimental to practical applications, and is in particular a problem with large compression factors, and this is exactly the challenge faced here where we seek to compress longer $>>$ 100 fs pulses to sub-20 fs duration.

Here we propose an efficient approach to improve few-cycle soliton compression by engineering multi-section structures in nonlinear crystals, such as quasi-phase-matching (QPM) structures with reduced residential phase mismatch or bonded $\beta$-barium borate (BBO) crystals with different orientation of optical axes, as shown in Fig. \ref{fig1}. The effective phase mismatch ($\Delta {k_{\rm{eff}}} $) in consecutive sections of nonlinear crystals decreases and always keeps $\Delta {k_{\rm{eff}}} > 0$ to make the cascading nonlinearity self-defocusing, and in this way the cascaded quadratic nonlinearity is engineered through the phase mismatch, a property which is not possible in standard Kerr media. The basic idea is that each section is designed to make the effective soliton order ${N_{\rm{eff}}} > 1$ by adjusting the residual (or QPM) phase mismatch, and so that at the exit of each section soliton compression is achieved. Such a compressed soliton will at this stage be in balance, so no further compression occurs unless we change the conditions, which is exactly what we propose. When this compressed pulse enters the new section the nonlinearity is engineered to make further compression occur. At the same time, each section is designed to keep a low effective soliton number, as soliton compression of pulses with small soliton numbers can achieve a small pedestal and high pulse quality. The numerical results turn out to work remarkably well already as a two-section and three-section design.

\section{Propagation equations and nonlinear Schr\textbf{$\ddot{\rm o}$}dinger equation in the cascading limit }

The temporal dynamics of ultra-short pulses propagating in the nonlinear media, specifically BBO and LN, can be described by the coupled wave equations under the slowly varying envelope approximation \cite{BacheJOSAB2007,BacheOE2008}:

\begin{align}
\label{eq1}
(i\frac{\partial }{{\partial z}} + {{\hat D}_1}){E_1} + {\rho _1}(z)E_1^*{E_2}{{\mathop{\rm e}\nolimits} ^{i\Delta {k_0}z}} + {\sigma _1}\left[ {{{\left| {{E_1}} \right|}^2}{E_1} + \eta {{\left| {{E_2}} \right|}^2}{E_1}} \right] = 0\\
(i\frac{\partial }{{\partial z}} - i{d_{12}}\frac{\partial }{{\partial t}} + {{\hat D}_{2}}){E_2} + {\rho _2}(z){E_1}^2{{\mathop{\rm e}\nolimits} ^{ - i\Delta {k_0}z}} + {\sigma _2}\left[ {{{\left| {{E_2}} \right|}^2}{E_2} + \eta {{\left| {{E_1}} \right|}^2}{E_2}} \right] = 0
\label{eq2}
\end{align}
where ${E_j}(z,t)$ is the complex amplitude of the electric field, ${\rho _j}$ and ${\sigma _j}$ involve quadratic and cubic nonlinearities, respectively, as ${\rho _j}(z) = {\omega _1}{d_{\rm{eff}}}(z)/c{n_j}$ and ${\sigma _j} = 3{\omega _j}{\mathop{\rm Re}\nolimits} ({\chi ^{(3)}})/8c{n_j}$. ${d_{{\rm{eff}}}}(z)$ is the quadratic nonlinear susceptibility along the propagation. The subscripts 1 and 2 correspond to the fundamental wave (FW) and second harmonic (SH) pulses. Here the phase mismatch $\Delta {k_0} = {k_2} - 2{k_1}$ and the material dispersion coefficients $k_j^{(n)} = {{{\partial ^n}{k_j}} \mathord{\left/ {\vphantom {{{\partial ^n}{k_j}} {\partial {\omega ^n}}}} \right. \kern-\nulldelimiterspace} {\partial {\omega ^n}}}\left| {_{\omega  = {\omega _j}}} \right.$, where ${k_j} = {{{n_j}{\omega _j}} \mathord{\left/
 {\vphantom {{{n_j}{\omega _j}} c}} \right. \kern-\nulldelimiterspace} c}$ is the wave numbers and ${n_j}$ is the refractive indices.
 Time $t$ is measured in a frame of reference travelling with the group velocity of the FW pulse.
 The group velocity mismatch (GVM) is ${d_{12}} = k_1^{(1)} - k_2^{(1)} = v_{g,1}^{ - 1} - v_{g,2}^{ - 1}$ and ${\hat D_j} = \sum\nolimits_{m = 2}^\infty  {\frac{{{i^m}}}{{m!}}k_j^{(m)}\frac{{{\partial ^m}}}{{\partial {t^m}}}} $ is total dispersion operator, which in frequency domain can be described as ${\hat D_j}(\Omega ) = k_j(\omega_j + \Omega) - \Omega k_1^{(1)} - k_j(\omega_j)$. The cross-phase modulation (XPM) coefficient is $\eta  = 2$ (or $2/3$) for type-0 (or type-I) SHG geometry.

Under the conditions of the cascading limit ($\Delta kL \gg 1$) and the negligible cross-phase-modulation (XPM) contribution, the soliton compression based on cascaded phase mismatched SHG process can be easily understood by reducing Eqs. (\ref{eq1}-\ref{eq2}) to a normalized nonlinear Schr$\ddot{\rm o}$dinger equation (NLSE) for the FW field ${U_1}$ (see details in Ref. \cite{BacheOE2008}):

\begin{equation}
(i\frac{\partial }{{\partial \xi }} + {\frac{1}{2}}\frac{{{\partial ^2}}}{{\partial {\tau ^2}}}){U_1} - (N_{{\rm{casc}}}^2 - N_{{\rm{Kerr}}}^2){U_1}{\left| {{U_1}} \right|^2} = iN_{{\rm{SHG}}}^2\frac{{2\delta }}{{\Delta \beta }}{\left| {{U_1}} \right|^2}\frac{{\partial {U_1}(\tau )}}{{\partial \tau }}
\label{eq3}
\end{equation}

The dimensionless field ${U_1}$ is normalized by the peak of the input electric field ${E_{\rm in}}$ and the time. The propagating distance are rescaled as $\tau  = (t - k_1^{(1)}z)/{T_0}$ and $\xi  = z/{L_d}$, where ${L_d} = T_0^2/| {k_1^{(2)}} |$ and ${T_0}$ are the FW dispersion length and input pulse duration. The dimensionless phase mismatch and GVM is $\Delta \beta  = \Delta {k_0}{L_d}$ and $\delta  = {d_{12}}{T_0}/| {k_1^{(2)}} |$.
Here we assume that $\Delta k_{\rm{eff}} > 0$ so the cascading nonlinearity is self-defocusing, and that the FW GVD is positive (normal dispersion). The effective soliton number is defined as $N_{{\rm{eff}}}^2 = N_{{\rm{casc}}}^2 - N_{{\rm{Kerr}}}^2 = {L_d}{\omega _1}{I_{\rm 0}}( {| {n_{{\rm{casc}}}^I} | - n_{{\rm{Kerr}}}^I} )/c$, where $I_{\rm 0}$ is input intensity and ${N_{{\rm{casc}}}}$ and ${N_{{\rm{Kerr}}}}$ are the dimensionless quadratic and cubic soliton numbers and $n_{{\rm{casc}}}^I \propto  - d_{{\rm{eff}}}^2/\Delta {k_{{\rm{eff}}}}$ is the Kerr-like cascading and ${n_{{\rm{Kerr}}}^I}$ is the self-focusing material Kerr nonlinearities, respectively.

Equation (\ref{eq3}) shows that with cascaded SHG, solitons can be excited in much the same way as in a Kerr medium \cite{BacheJOSAB2007}: In order for few-cycle pulse compression to occur based on phase-mismatched SHG, an effective self-defocusing nonlinearity ${N_{\rm{eff}}} >$ 1 with normal dispersion is required. In the same way as for solitons in a pure Kerr medium, pulse compression with higher-order solitons suffer from low pulse quality and a large unwanted uncompressed pedestal. The right-hand side of Eq. (\ref{eq3}) contains the first-order contribution (in the weakly nonlocal approximation) of the cascading response. It is a GVM-induced contribution, which acts in a similar way as self-steepening \cite{MosesPRL2006}.

\section{Numerical results and discussions}

To obtain accurate numerical simulation of few-cycle pulse region, we use the full coupled SHG propagation equations under the slowly-evolving-wave approximation including exact dispersion, self-steepening and electronic Kerr effects \cite{BacheOE2008} (see \cite{MosesPRL2006} for more details). We consider 5$\%$ MgO doped periodically poled lithium niobate (PPLN: MgO) for pulse compression at the wavelength of 1.56 $\mu{\rm m}$, where high-energy Er-doped fiber amplifiers operate. Type-0 phase-mismatched SHG is used for soliton compression because of its large quadratic nonlinear component (at 1520 nm the value ${d_{{\rm{33}}}}$  = 20.6 pm/V was measured recently \cite{Schiek2012} and the electronic Kerr nonlinear refractive index is chosen to be ${n^I}_{{\rm{Kerr}}} = 3{\mathop{\rm Re}\nolimits} ({\chi ^{(3)}})/4n_1^2{\varepsilon _0}c = 23 \cdot {10^{ - 20}}{{{{\rm{m}}^{\rm{2}}}} \mathord{\left/
 {\vphantom {{{{\rm{m}}^{\rm{2}}}} {\rm{W}}}} \right.
 \kern-\nulldelimiterspace} {\rm{W}}}$ \cite{ZhouPRL}). The refractive indices and the material dispersion coefficients are derived from Sellmeier's equation for MgO: LN \cite{Gayer2008}. The balancing point between quadratic and cubic nonlinearities where $\left| {n_{{\rm{casc}}}^I} \right| = n_{{\rm{Kerr}}}^I$ determines the upper limit of the effective phase mismatch for soliton formation in normal dispersion, which is calculated to be less than 212${~{\rm mm^{-1}}}$ in QPM and the effective phase mismatch is $\Delta {k_{j,{\rm{QPM}}}} = \Delta {k_0} - {{2\pi } \mathord{\left/
 {\vphantom {{2\pi } {{\Lambda _j}}}} \right.
 \kern-\nulldelimiterspace} {{\Lambda _j}}}$  and  ${\Lambda _j}$ is the QPM period. Therefore we always choose $\Delta {k_{1,{\rm{QPM}}}}$ in the first QPM section to be 180${~{\rm mm^{-1}}}$ in the following simulations of soliton compression in multi-section QPM gratings.

\begin{figure}[htbp]
  \centering{

  \includegraphics[width=10cm]{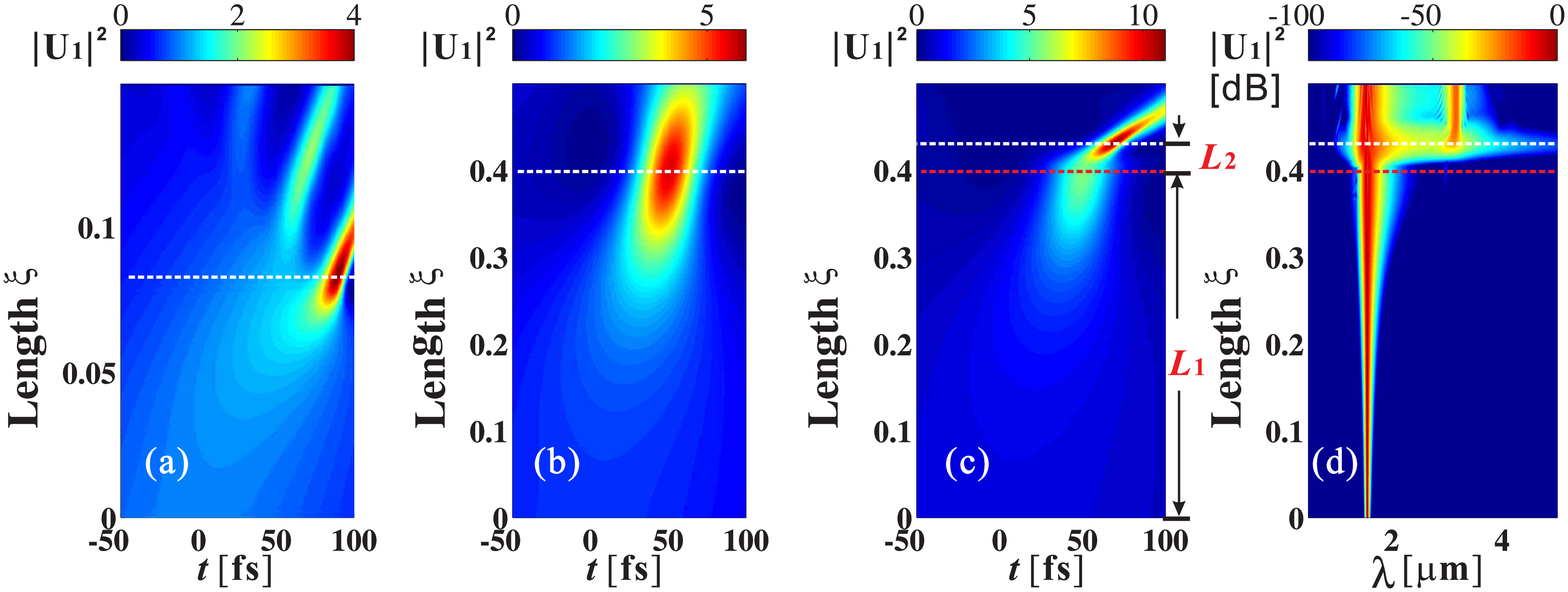}
  \includegraphics[width=12cm]{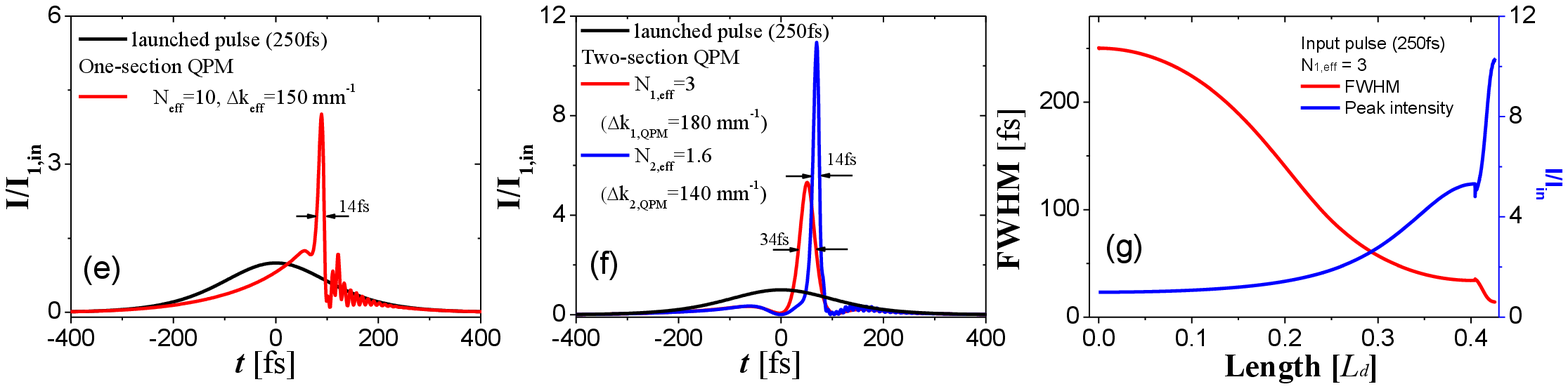}

 }
\vspace{-2mm}

\caption{Numerical simulation of compressed pulses of ${T_{\rm in}}$ = 250 fs FWHM at the pump wavelength of 1.56$~\mu{\rm m}$ in one- and two-section QPM structures. FW temporal evolutions of (a) ${N_{{\rm{1,eff}}}} = 10$, $\Delta {k_{{\rm{QPM}}}}$ = 150${~{\rm mm^{-1}}}$ and (b) ${N_{{\rm{1,eff}}}} = 3$, $\Delta {k_{{\rm{QPM}}}}$ = 180${~{\rm mm^{-1}}}$ in one-section QPM structure; (c) and (d) FW temporal and spectral evolution in two-section QPM structure with ${N_{{\rm{1,eff}}}} = 3$, $\Delta {k_{{\rm{1,QPM}}}}$ = 180${~{\rm mm^{-1}}}$ (${N_{{\rm{2,eff}}}} = 1.6$, $\Delta {k_{{\rm{2,QPM}}}}$ = 140${~{\rm mm^{-1}}}$) in the first (second) QPM. (e) and (f) Normalized intensities of the input, the optimal compression positions [white dashed lines, cuts in (a), (b) and (c)]. (g) Pulse durations and the ratio of peak intensities along propagation in two-section QPM.}
\label{fig2}
\end{figure}

The input transform-limited FF pulse duration is 250 fs at full width at half-maximum (FWHM) at the pump wavelength of 1.56$~\mu{\rm m}$. Figure \ref{fig2}(a)-(b) show the numerical simulations of soliton compression with different effective soliton numbers (${N_{{\rm{1,eff}}}}$) in single-section QPM structures. The pulse quality of the compressed FW is well maintained by keeping a low soliton order but the drawback is a limited smaller compression factor. Figure \ref{fig2}(c)-(d) shows the FW temporal and spectral evolution in a two-section QPM structure. The effective soliton number in the first QPM section ${N_{{\rm{1,eff}}}}$ is chosen to 3, corresponding to a peak intensity of 25$~{\rm GW/cm^2}$ when $\Delta {k_{{\rm{1,QPM}}}}$ = 180${~{\rm mm^{-1}}}$. The temporal propagation dynamics for such a configuration is shown in Fig. \ref{fig2}(b). The compressed pulse reaches its minimal duration of 34 fs after propagating the distance of ${L_1}$ = 0.4$~{L_{1,d}}$ (the dispersion length is ${L_{1,d}}$  = 211 mm in the first QPM section) and continues to compress to 14 fs in the second QPM section, which has $\Delta {k_{{\rm{2,QPM}}}}$ = 140${~{\rm mm^{-1}}}$, which corresponds to ${N_{{\rm{2,eff}}}}$ = 1.6 for a 34 fs pulse. In the spectrum, strong broadening of the FW is observed, and the soliton formation also leads to a soliton-induced optical Cherenkov wave around the wavelength of 3$~\mu{\rm m}$, as shown in Fig. \ref{fig2}(d). The FW pulse has a slight oscillation upon propagation in the onset of the second QPM and then continues to compress further and the peak intensity increases quickly, as shown in Fig. \ref{fig2}(f).

Note that the dispersion length in the $j$th-section QPM is defined as  ${L_{j,d}} \equiv {{T_{j,{\rm in}}^2} \mathord{\left/
 {\vphantom {{T_{j,{\rm in}}^2} {|k_1^{(2)}|}}} \right.
 \kern-\nulldelimiterspace} {|k_1^{(2)}|}}$ and $T_{j,{\rm in}}$ is the pulse duration at the input of $j$th-section QPM. The second distance ${L_2}$ for optimal compression is much shorter than ${L_1}$, but it corresponds to $1.4~{L_{2,d}}$ and the dispersion length ${L_{2,d}}$ is 4 ${\rm mm}$. The quality factor ($Q_c$) of the compressed pulse, defined as the ratio of the fractional amount of energy carried by the central spike of the FW pulse over the input pulse energy, is as high as 0.63. The pulse compression of higher-order soliton compression (${N_{{\rm{eff}}}} = 10$) in a single-section QPM is also shown in Fig. \ref{fig2}(e) for comparison ($Q_c$ is 0.22). The pulse quality is much worse than with two-section QPM: a large pedestal and trailing oscillations are observed, even though the main spike has pulse duration of 14 fs. The length required for multi-section QPM is in the example shown quite long (around 85 ${\rm mm}$), but it can be reduced by increasing the soliton number. For instance, the total length required for pulse compression of the input FF pulse (250 fs) with the soliton number ${N_{{\rm{1,eff}}}} = 5$ is 47 ${\rm mm}$. This comes at a price, though, as the quality factor is reduced to 0.45 and the input intensity increases. Note also if LN is used at shorter wavelengths, the dispersion length becomes shorter due to the increase of material GVD. On the other hand, input Gaussian beam with a longer focus length than the device length is required and leads to the reduced pump intensity. Fortunately, our compression scheme is accomplished under a low soliton order, in which the pump intensity is low. Hence a laser pulse with an energy on the order of ten micro-Joules can meet the requirement.

\begin{figure}[htbp]
  \centering{

\includegraphics[width=11cm]{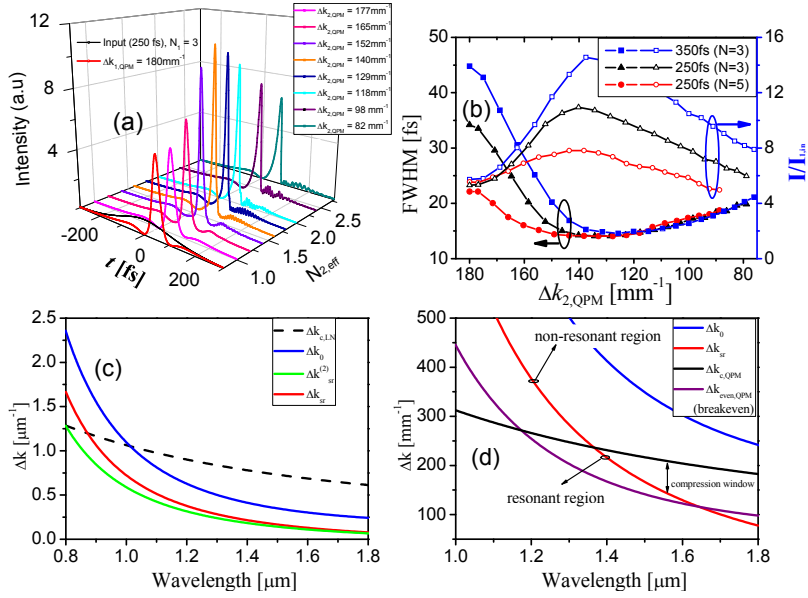}

}
\vspace{-2mm}
\caption{(a) Optimal compressed pulse intensities for selected values of the effective phase mismatch  $\Delta {k_{{\rm{2,QPM}}}}$ of the second QPM and (b) pulse durations and peak intensities versus  $\Delta {k_{{\rm{2,QPM}}}}$. (c) Wavelength dependence of $\Delta {k_{{\rm{0}}}}$ from material dispersion, the resonant threshold $\Delta {k^{(2)}_{{\rm{sr}}}}$, the upper limit of $\Delta {k_{{\rm{c,LN}}}}$ for soliton in unpoled LN. (d) Compression windows versus the wavelength between the upper limit of $\Delta {k_{{\rm{c,QPM}}}}$ (black line), the resonant threshold $\Delta {k_{{\rm{sr}}}}$ (red line) including full SH dispersion compression and the breakeven QPM value $\Delta {k_{{\rm{even,QPM}}}}$ (purple line) to achieve the same $n_{{\rm{casc}}}^I$ in unpoled LN.}
\label{fig3}
\end{figure}

Figure \ref{fig3}(a) shows the simulations of soliton compression by changing the effective wave mismatch $\Delta {k_{{\rm{2,QPM}}}}$ of the second QPM section. The optimal pulse durations and peak intensities for different input pulse durations and soliton numbers are shown in Fig. \ref{fig3}(b). It is clearly seen that all of the best compression cases (i.e. pulses with the highest intensity and shortest duration) occur in second-section QPM when $\Delta {k_{{\rm{2,QPM}}}}$ approaches to 140${~{\rm mm^{-1}}}$, at which the critical transition occurs from the nonstationary to stationary regions in the nonlocal model \cite{BacheOL2007}, i.e., the nonlocal response is localized in the stationary regime and oscillating in the nonstationary region.

The detailed discussions on the transition to the nonstationary regimes of BBO crystal were presented based on the nonlocal theory \cite{BacheOL2007,BacheJOSAB2007,BacheOE2008}
. Here we show the nonlocal analysis of stationary and nonstationary regimes in type-0 phase-mismatching SHG geometry of PPLN: MgO in the normal dispersion regime. Now including the effective phase mismatch in QPM, the nonlocal response function in frequency domain is derived as

\begin{equation}
R(\Omega ) \equiv {{\Delta {k_{{\rm{QPM}}}}} \mathord{\left/
 {\vphantom {{\Delta {k_{{\rm{QPM}}}}} {\Delta {k_{nl}}(\Omega )}}} \right.
 \kern-\nulldelimiterspace} {\Delta {k_{nl}}(\Omega )}}
 \label{eq4}
\end{equation}
and
\begin{equation}
\Delta {k_{nl}}(\Omega ) = {\hat D_2}(\Omega ) - {d_{12}}\Omega  + \Delta {k_{{\rm{QPM}}}} = k_2^{}(\omega {}_2 + \Omega ) - 2k_1^{}(\omega {}_1) - \frac{{2\pi }}{\Lambda } - k_1^{(1)}\Omega
\label{eq5}
\end{equation}
It is seen that when $\Delta {k_{{\rm{QPM}}}}$ is below a certain threshold $\Delta {k_{{\rm{QPM}}}} < \Delta {k_{{\rm{sr}}}}$, $R(\Omega )$ becomes resonant spectrally and consequently the SH spectrum develops a characteristic sharp resonance peak (in the time domain the SH pulse becomes stretched and can eventually span many picoseconds)\cite{Bache2010PRA}. This is of course an unwanted property for efficient femtosecond pulse interaction, where a large bandwidth is required; when $\Delta {k_{{\rm{QPM}}}}$ is significantly larger than the threshold $\Delta {k_{{\rm{QPM}}}} \ge \Delta {k_{{\rm{sr}}}}$, $R(\Omega )$ is non-resonant and holds octave-spanning bandwidths \cite{ZhouPRL}. Therefore $\Delta {k_{{\rm{QPM}}}} \ge \Delta {k_{{\rm{sr}}}}$ indicates the boundary to the stationary regime, where ultra-broadband non-resonant interaction is supported. If only including up to second-order SH dispersion, we can explicitly obtain the threshold $\Delta k_{{\rm{sr}}}^{(2)} = d_{12}^2/2k_2^{(2)}$ and by using full SH dispersion, $\Delta k_{{\rm{sr}}}^{} \approx $ 143${~{\rm mm^{-1}}}$, which is quite close to the numerical results of the phase-mismatch value of the optimal compression in the second QPM gratings.

QPM reduces the effective quadratic nonlinearity $d_{\rm eff}$ (with at least 2/$\pi$ ) and the breakeven phase-mismatch value $\Delta k_{{\rm{even,QPM}}} = 4{{\Delta {k_0}} \mathord{\left/
 {\vphantom {{\Delta {k_0}} {{\pi ^2}}}} \right.
 \kern-\nulldelimiterspace} {{\pi ^2}}} \approx $ 130${~{\rm mm^{-1}}}$ to achieve the same nonlinearity as in unpoled LN ($n_{{\rm{casc}}}^I(\Delta {k_0}) = n_{{\rm{casc}}}^I(\Delta k_{{\rm{even,QPM}}}) = 38 \cdot {10^{ - 20}}{{{{\rm{m}}^{\rm{2}}}} \mathord{\left/
 {\vphantom {{{{\rm{m}}^{\rm{2}}}} {\rm{W}}}} \right.
 \kern-\nulldelimiterspace} {\rm{W}}}$). This phase mismatch value turns out to be in the non-stationary regime, while the inherent (material) phase mismatch $\Delta {k_0}$ is large enough to be in the stationary regime and still be below the upper limit of $\Delta {k_{{\rm{c,LN}}}}$ for soliton compression, as shown in Fig. \ref{fig3}(c). The cascading nonlinearity using QPM when $\Delta {k_{{\rm{QPM}}}} > \Delta {k_{{\rm{sr}}}}$, is therefore smaller than that in unpoled LN, and this property exhibits the advantage by using noncritical SHG for soliton compression in unpoled LN over QPM LN \cite{ZhouPRL}. However in the multi-section structure, unpoled LN cannot be used for the first-stage compression. As in the following QPM section $n_{\rm{casc}}^I(\Delta k_{\rm{2,QPM}})$ need to be increased, there is no QPM phase-mismatch space for second-stage compression because in the second stage, the cascaded nonlinear index needs to be larger than that in the first stage, seen in Fig. \ref{fig1}. One reason is that the breakeven phase mismatch value $\Delta k_{\rm{even,QPM}}$ in unpoled LN is close to the non-stationary regime and another is that the larger cascaded nonlinearity of second section PPLN requires smaller phase mismatching value because QPM modulation gives a pre-factor ($2/\pi$ from the first order).

 QPM can engineer $\left| {n_{{\rm{casc}}}^I} \right| \propto d_{{\rm{eff}}}^2/\Delta {k_{{\rm{QPM}}}}$ by changing the residual phase mismatch with different QPM domain lengths, even if this means sacrificing the cascading strength. Taking into account that optimal soliton compression occurs in the stationary regime ($\Delta {k_{\rm{eff}}} > \Delta {k_{\rm{sr}}}$) and that there is an upper limit of the effective phase mismatch for soliton formation in normal dispersion, we obtain the compression window of $\Delta {k_{{\rm{QPM}}}}$  in type-0 phase-matching SHG geometry of PPLN: MgO, as shown in Fig. \ref{fig3}(d). This shows enough $\Delta {k_{{\rm{QPM}}}}$ space for fully exploiting quadratic nonlinearities by engineering the QPM structure, for example,  $n_{{\rm{casc}}}^I(\Delta {k_{{\rm{QPM}}}} = 180{\rm{mm}}^{-1}) = 27 \cdot {10^{-20}}{\rm{m}^2}/{\rm{W}}$ and $n_{{\rm{casc}}}^I(\Delta {k_{{\rm{QPM}}}} = 140{\rm{mm}}^{- 1}) = 35 \cdot {10^{ - 20}}{\rm{m}^2}/{\rm{W}}$. It is noted that the critical wavelength where the compression window opens is around 1.37 $\mu{\rm m}$ due to the reduced cascading nonlinearity, which longer than that in unpoled LN, where the window opens already at 0.9 $\mu{\rm m}$.

\begin{figure}[htbp]
  \centering{

\includegraphics[width=12cm]{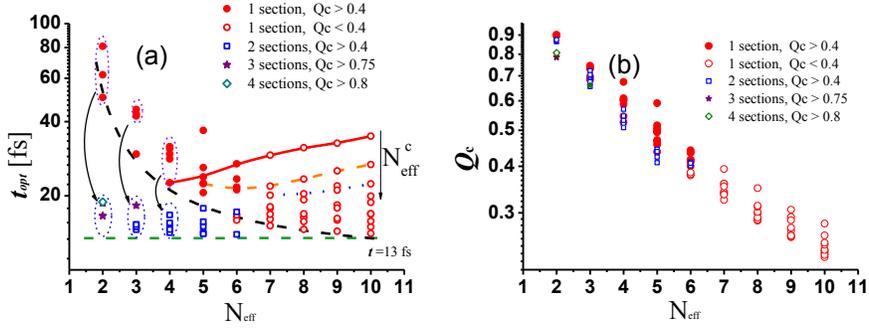}

}
\vspace{-2mm}
\caption{Results of numerical simulations showing the optimum compression parameters versus ${N_{{\rm{eff}}}}$ in multi-section QPM gratings for (a) the minimum pulse duration, (b) quality factor. The filled mark with round symbols are the compression results in one section QPM with good quality factor ($Q_c > 0.4$) and the square, star and diamond symbols represents the results in multi-section QPM (two, three and four) with high quality factor while durations are close to the empirical limit of 13 fs (green dash line). In (a), the arrow shows stronger pulse compression with the decrease of $N_{{\rm{eff}}}^{\rm{c}}$ and the black dash line is fit to the minimum duration available in one-section QPM gratings. The green dashed line in (a) denotes an empirical limit of 13 fs that was observed in the simulations.}
\label{fig4}
\end{figure}

  Figure \ref{fig4}(a)-(b) compare in one- and multi-section QPM design the quality factor and the optimum pulse duration versus the effective soliton number. In single-section QPM, a high soliton order is required to compress the pulse towards few-cycle duration, but this leads to much lower $Q_c$; the compressed pulses with duration of an empirical limit of 13 fs and good quality factors ($Q_c > 0.4$) are achieved in two-section QPM by using the input pulses of small soliton numbers (${N_{{\rm{eff}}}}$ = 3, 4 and 5). The minimal pulse durations of 18 fs with a high quality factor 0.8 are achieved in three- and four-section QPM, and evidently the trend is that many sections and low soliton orders are favorable. The caveat is that the total propagation length becomes very long, which makes it difficult to be realized practically in a single QPM crystal, so as always one must find a good compromise in a final design.

   In fact, input pulses with a large soliton number will experience a stronger competing Kerr XPM effect, which is also self-focusing and reduces the overall self-defocusing nonlinearity. The effective soliton number corrected by XPM effect can be expressed as
   \begin{equation}
   N_{{\rm{eff(XPM)}}}^2 = N_{{\rm{casc}}}^2 - N_{{\rm{Kerr(SPM)}}}^2 - N_{{\rm{Kerr(XPM)}}}^2 = N_{{\rm{eff}}}^2 - N_{{\rm{Kerr(XPM)}}}^2
   \label{eq6}
\end{equation}
and
 \begin{equation}
N_{{\rm{Kerr(XPM)}}}^2 = N_{{\rm{Kerr}}}^2(1 + B{{{n_1}N_{\rm casc}^2} \mathord{\left/
 {\vphantom {{{n_1}N_{\rm casc}^2} {{n_2}\Delta {k_{{\rm{eff}}}}}}} \right.
 \kern-\nulldelimiterspace} {{n_2}\Delta {k_{{\rm{eff}}}}}})
 \label{eq7}
\end{equation}

 Thus the critical value of the effective soliton number for soliton behavior requires
 \begin{equation}
 N_{{\rm{eff}}}^{} \ge N_{{\rm{eff}}}^{\rm{c}} = \sqrt {1 + {{\Phi _{{\rm{in}}}^2} \mathord{\left/
 {\vphantom {{\Phi _{{\rm{in}}}^2} {\Phi _{\rm{c}}^2}}} \right.
 \kern-\nulldelimiterspace} {\Phi _{\rm{c}}^2}}}
 \label{eq8}
 \end{equation}
  where
  $\Phi _{\rm{c}}^{} \equiv \sqrt {{{4\left| {k_1^{(2)}\Delta k} \right|{n_2}{c^2}} \mathord{\left/
 {\vphantom {{4\left| {k_1^{(2)}\Delta k} \right|{n_2}{c^2}} {(B{n_{{\rm{Kerr}}}}\left| {{n_{{\rm{casc}}}}} \right|{n_1}\omega _1^2)}}} \right.
 \kern-\nulldelimiterspace} {(B{n_{{\rm{Kerr}}}}\left| {{n_{{\rm{casc}}}}} \right|{n_1}\omega _1^2)}}} $
 . $\Phi _{{\rm{in}}}$ is the energy fluence of the input pulse, ${\Phi _{\rm in}} = 2{T_{\rm in}}{I_{\rm in}}$ for a ${\rm sech}^2$-shaped pulse. $N_{{\rm{eff}}}^{\rm{c}}$ is clearly dependent on the input fluence and the phase mismatch. In Fig. \ref{fig4}(a), we show that by using large soliton number stronger pulse compression occurs as $N_{{\rm{eff}}}^{\rm{c}}$ decreases.

\begin{figure}[htbp]
  \centering{
  \includegraphics[width=10cm]{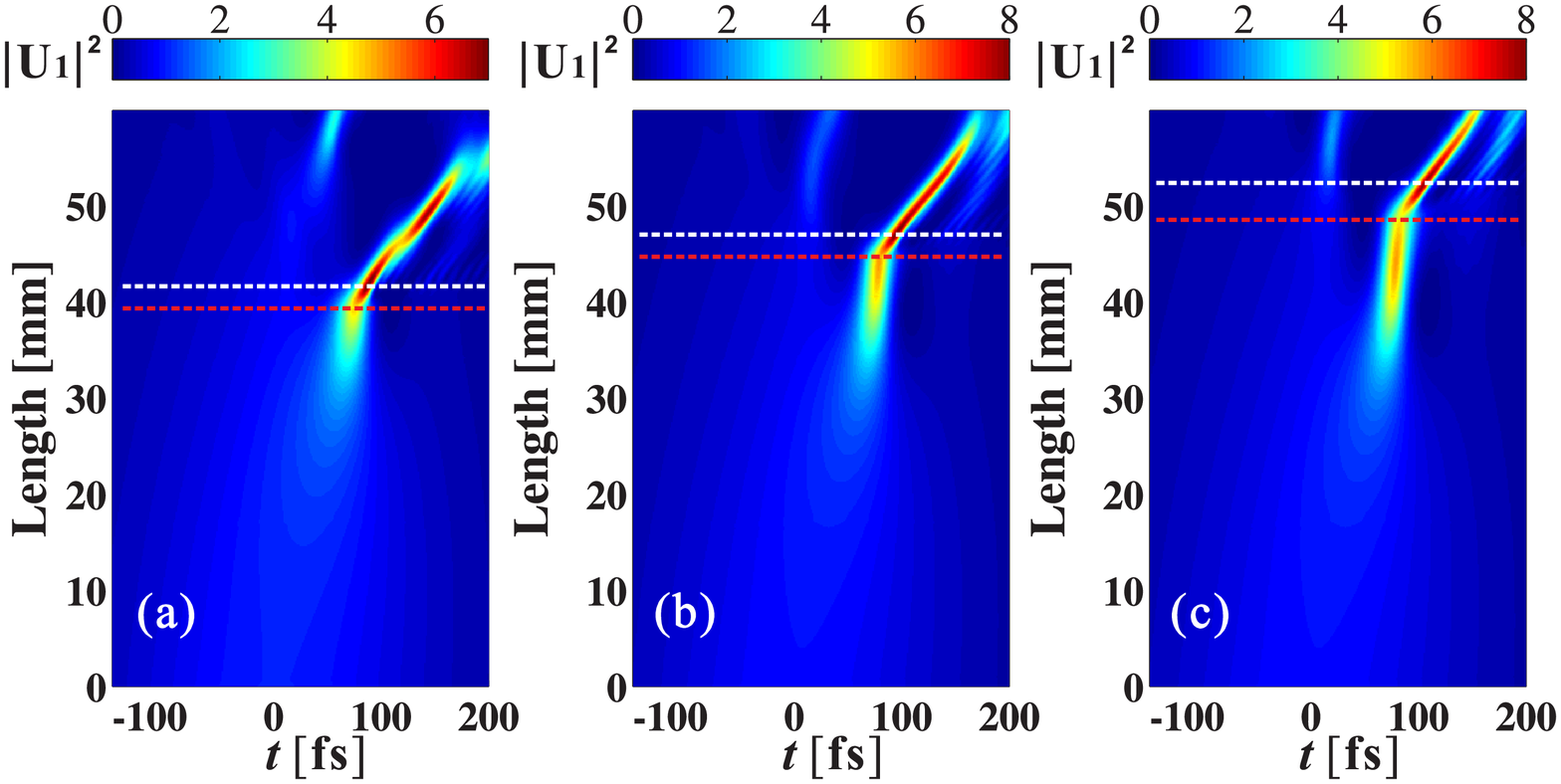}
  \includegraphics[width=10cm]{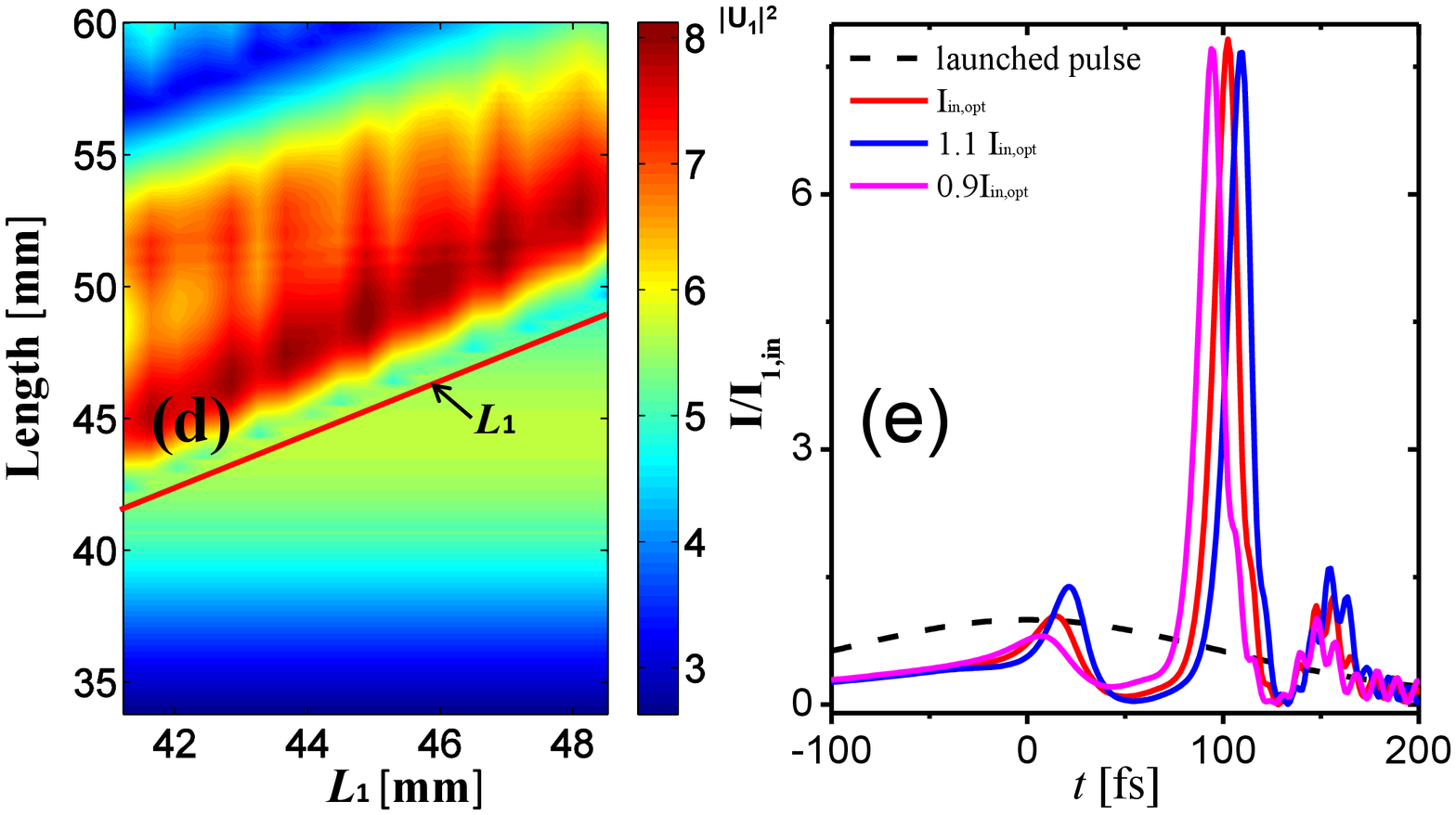}
}

\vspace{-2mm}
\caption{Temporal evolutions of FW pulse (250 fs FWHM) in two-section QPM structures ($N_{\rm {1,eff}}$  = 5): the length of first-section QPM is (a) 39.2 mm (b) 44.5 mm (optimal compression position) and (c) 48.5 {\rm mm}. (d) Peak intensifies of FW along propagation as function of the length of first QPM. (e) Comparison of the final output pulses with 10$\%$ fluctuation of input intensities in same QPM structure ($L_1$ = 44.5 mm and $L_2$ = 3 mm).}
\label{fig5}
\end{figure}
One may question whether the approach works if the input parameters are different from the pre-designed values. The numerical results confirm the capability of generating high-quality compressed pulses, as shown in Fig. \ref{fig5}(a)-(c), when soliton compression in the first-stage QPM occurs early or later, which is beneficial from low-order soliton propagation by controlling quadratic nonlinearities. Figure \ref{fig5}(d) plots the peak intensifies of FW along the propagation direction as function of the first QPM length $L_1$. Soliton compression to few-cycle pulses still occur in second-section QPM even when $L_1$ has become shorter or longer than the optimal length. Similarly, the capability to survive some fluctuation of input pulses is also seen in Fig. \ref{fig5}(e). The final output pulses have slight degradation when 10$\%$ fluctuations of pulse intensity are inputted in the same QPM structure ($L_1$ = 44.5 mm and $L_2$ = 3 mm).

Recently Raman nonlinear contribution of lithium niobate in cascaded nonlinear interaction have been carefully considered in supercontinuum generation \cite{Phillips2011}, few-cycle soliton compression and optical Cherenkov radiation \cite{Bache2010PRA,Bache2011OE,ZhouPRL}. In the time scale of few-cycle pulses, Raman Kerr response of lithium niobate becomes quite complicated and relies on the property of material composition
 \cite{BacheNote}. In the present study, Raman fractions $f_R$ = 0.14 (0.50) are used in the simulations, while keeping the same electronic Kerr contributions.

\begin{figure}[htbp]
  \centering{

  \includegraphics[width=12cm]{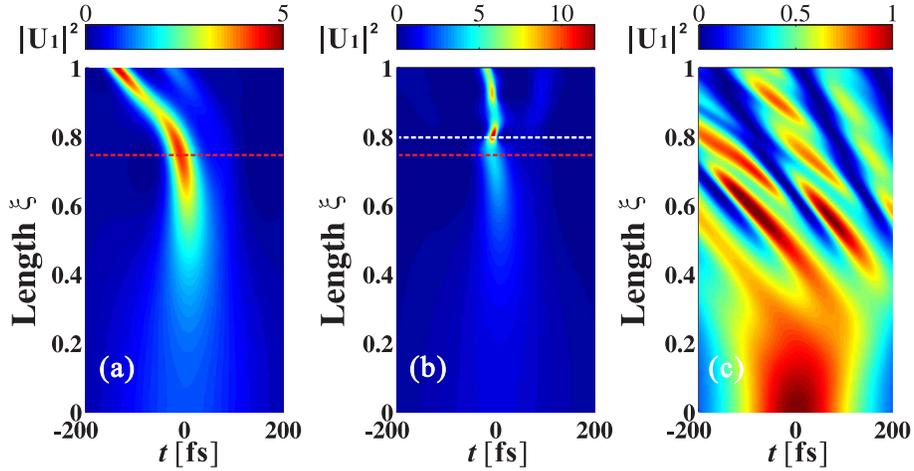}
}
\vspace{-2mm}
\caption{Numerical results of temporal evolution of FW pulse (250 fs FWHM) including Raman contribution, (a) one-section (b) two-section QPM when $f_R$ = 0.14 and (c) one-section QPM when $f_R$ = 0.50.}
\label{fig6}
\end{figure}

In the case of small Raman fraction $f_R$ = 0.14, Fig. \ref{fig6}(a) shows a simulation in a single-section QPM structure. Compared with Fig. \ref{fig2}(b) where Raman was not included, the simulation shows that the pulse is sped up instead of being delayed. This is because in the normal-dispersion regime, the soliton that is becoming red-shifted spectrally by the Raman effect will travel faster and it even overcomes the initial pulse delay from the GVM-induced self-steepening (right-hand side of Eq. (\ref{eq3})). For this value of the Raman fraction we can still implement the soliton compression to achieve high-quality few-cycle pulses by using multi-section QPM structure, as shown in Fig. \ref{fig6}(b).  As the design imposes a decrease of $\Delta {k_{{\rm{QPM}}}}$ in the second-section QPM, the GVM-induced self-steepening increases (as it is proportional to $1/\Delta{k_{\rm{QPM}}}$) to reduce the overall Raman influence and even leads to possible balance; this will discussed further in a separate publication \cite{Guo2012}. However, when the Raman fraction is increased to $f_R$ = 0.50, a sort of modulational instability occurs as seen in Fig. \ref{fig6}(c). Note that it seems not to be soliton fission, which usually happens after the first compression point, as it seems that the pulse does not experience complete compression before pulse splitting happens. Therefore the Raman response of the nonlinear materials plays important role in the cascaded soliton compression and the dynamics will be published elsewhere.

\begin{figure}[htbp]
  \centering{

  \includegraphics[width=12cm]{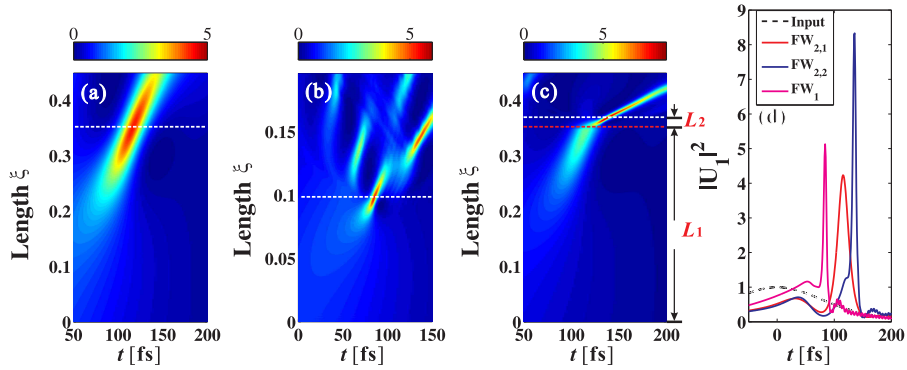}

}
\vspace{-2mm}
\caption{Evolution of the temporal profiles of FW pulses (T$_{\rm in}$ = 200 fs FWHM) in BBO crystals. Propagation in one section (a) $N_{\rm eff}$  = 3.5,  $\Delta {k_{{\rm{eff}}}}$ = 75 mm${^{-1}}$ (b) $N_{\rm eff}$  = 8,  $\Delta {k_{{\rm{eff}}}}$ = 60 mm${^{-1}}$; (c) $\Delta {k_{{\rm{1}}}}$ = 75 mm${^{-1}}$ ($N_{\rm1,eff}$  = 3.5) and $\Delta {k_{{\rm{2}}}}$ = 58 mm${^{-1}}$ in two-section BBO crystal
($L_1$ = 93.9 mm and $L_2$ = 4.5 mm); (d) pulse intensities at the optimal compression positions (dashed lines) in one-section (FW$_1$) and two-section (FW$_{2,1}$: first compression; FW$_{2,2}$: second compression) BBO crystals.}
\label{fig7}
\end{figure}

Finally we consider BBO as the nonlinear medium in type-I birefringent phase-matching SHG geometry, in which Raman contribution is negligible under the relevant experimental conditions. BBO has a decent quadratic nonlinear strength ( ${d_{{\rm{eff}}}}$ = 2.06 pm/V) and a very low cubic nonlinear refractive index ( ${n^I}_{{\rm{Kerr}}} = 5.90 \cdot {10^{ - 20}}{\rm{m}^2}/{\rm{W}}$) at the wavelength of 0.8 $\mu{\rm m}$ \cite{BacheNoteBBO}. Compared to LN, the FW GVD in BBO crystal is smaller, which leads to longer dispersion lengths. Therefore we use in the simulations shorter durations (200 fs FWHM) of input pulse and the wavelength of 1.030 $\mu{\rm m}$, which is current state of the art for Yb-based solid-state amplifiers. As shown in Fig. \ref{fig1}(d), the phase mismatch can be altered by changing the angle $\theta$ between the FW input and the optical z axis of the crystal. A multi-section structure of BBO crystals with different angles $\theta$ may be constructed through some epoxy-free optical bonding techniques, as seen in \cite{Myatt2006}. Alternatively, two BBO crystals could be put in sequence (one long and one short) where individual angle-tuning of the phase-mismatch values gives a large operational flexibility to achieve optimal pulse compression. Achieving huge compression factors from initial 30-ps to 30 fs pulses was proposed based on multi-stage pulse compression by use of cascaded quadratic nonlinearity in BBO \cite{Xie2007OptComm}, where additional beam expanders are needed to reduce the peak intensities. The difference to the current work is that we always keep soliton compression in the stationary regimes to achieve octave-spanning bandwidths that are large enough for supporting few-cycle pulses and use small soliton numbers in each section compression to produce high-quality pulses.

The numerical results shown in Fig. \ref{fig7} confirm that pulse compression in two-section BBO crystals is greatly improved compared to one-single BBO crystal. Figure \ref{fig7}(a) shows a single-stage compression simulation with a low soliton order. Figure \ref{fig7}(b) shows that high-order soliton compression may reach the limit of pulse duration (6 fs) but leave most of energy in the pedestal ($Q_c$ = 0.23); while keeping low-order soliton compression in two-section BBO crystals can hold high-quality pulse ($Q_c$ = 0.72) with maximum peak intensity while minimum pulse width approaching the same duration as shown in Fig. \ref{fig7}(c). Figure \ref{fig7}(d) plots the pulse intensities at the optimal positions in one- and two-section BBO crystals.

\section{Conclusion}
We proposed a new scheme for compressing longer $>> 100$  fs pulses through a soliton effect to few-cycle duration with high quality. The cascaded quadratic nonlinearity was engineered through varying the phase mismatch in ways of multi-section  quasi-phase-matching (QPM) structure with different domain periods and bonded BBO crystals with different orientation of the optical axis. By adjusting the residual or QPM phase mismatch, we made ${N_{{\rm{eff}}}} > 1$ in the beginning of each section and kept soliton compression in the stationary regimes, so that clean few-cycle pulse compression was accessible. At the same time, each section was designed to keep a low effective soliton number, as soliton compression of pulses with small soliton numbers can achieve a small pedestal and high pulse quality.

We in detail presented the numerical results of soliton compression in multi-section QPM structure at the wavelength of 1.56 $\mu{\rm m}$. The available cascaded nonlinearities are compared in unpoled lithium niobate (LN) and periodically poled lithium niobate (PPLN), and we showed by using QPM the advantage of enough QPM phase-mismatch space for second-stage compression. However, the shortest available wavelength for pulse compression in the QPM scheme is around 1.37 $\mu{\rm m}$, due to competing Kerr nonlinearities. The wavelength cut off is substantially higher than that in LN, where compression may occur even down to around 1 micron. Examples of soliton compression to 13 fs (below three optical cycles) were summarized in multi-section QPM with different input pulse duration and soliton numbers. Input pulses with large soliton number experienced a stronger competing Kerr XPM effect, which is self-focusing and therefore reduced the overall self-defocusing nonlinearity.

Numerical results showed that the compressed pulse with less than three-cycle duration was close to the limit of the higher-order soliton compression while keeping quality factor more than 0.40. We also discussed that the tolerance of input intensities and the error of the QPM structures affect the soliton compression in the multi-section QPM. The Raman contribution to few-cycle soliton compression was also discussed.

Finally we implemented the scheme in bonded BBO crystals at the wavelength of 1.03 $\mu{\rm m}$. The numerical results confirmed that pulse quality of soliton compression in two-section BBO crystals is greatly improved while minimum pulse width approaching the same duration (6 fs) compared to one-single BBO crystal.

\section*{Acknowledgments}

 Xianglong Zeng acknowledges the support of Marie Curie International Incoming Fellowship from EU (grant No. PIIF-GA-2009--253289) and the financial support from National Natural Science Foundation of China (60978004) and Shanghai Shuguang Program (10SG38). Morten Bache acknowledges the support from the Danish Council for Independent Research (Technology and Production Sciences, grant No. 274-08-0479 "Femto-VINIR", and grant No. 11-106702 "Femto-midIR").

\end{document}